\input harvmac
\noblackbox
\newcount\figno
\figno=0
\def\fig#1#2#3{
\par\begingroup\parindent=0pt\leftskip=1cm\rightskip=1cm\parindent=0pt
\baselineskip=11pt
\global\advance\figno by 1
\midinsert
\epsfxsize=#3
\centerline{\epsfbox{#2}}
\vskip 12pt
\centerline{{\bf Figure \the\figno:} #1}\par
\endinsert\endgroup\par}
\def\figlabel#1{\xdef#1{\the\figno}}

\def\np#1#2#3{Nucl. Phys. {\bf B#1} (#2) #3}
\def\pl#1#2#3{Phys. Lett. {\bf B#1} (#2) #3}
\def\prl#1#2#3{Phys. Rev. Lett. {\bf #1} (#2) #3}
\def\prd#1#2#3{Phys. Rev. {\bf D#1} (#2) #3}

\def\cmp#1#2#3{Comm. Math. Phys. {\bf #1} (#2) #3}


\font\cmss=cmss10
\font\cmsss=cmss10 at 7pt
\def\rlx{\relax\leavevmode}
\def\inbar{\vrule height1.5ex width.4pt depth0pt}
\def\IC{\relax\,\hbox{$\inbar\kern-.3em{\rm C}$}}
\def\IN{\relax{\rm I\kern-.18em N}}
\def\IP{\relax{\rm I\kern-.18em P}}
\def\ZZ{\rlx\leavevmode\ifmmode\mathchoice{\hbox{\cmss Z\kern-.4em Z}}
 {\hbox{\cmss Z\kern-.4em Z}}{\lower.9pt\hbox{\cmsss Z\kern-.36em Z}}
 {\lower1.2pt\hbox{\cmsss Z\kern-.36em Z}}\else{\cmss Z\kern-.4em
 Z}\fi}
\def\IZ{\relax\ifmmode\mathchoice
{\hbox{\cmss Z\kern-.4em Z}}{\hbox{\cmss Z\kern-.4em Z}}
{\lower.9pt\hbox{\cmsss Z\kern-.4em Z}}
{\lower1.2pt\hbox{\cmsss Z\kern-.4em Z}}\else{\cmss Z\kern-.4em
Z}\fi}

\def\narrowplus{\kern -.04truein + \kern -.03truein}
\def\narrowminus{- \kern -.04truein}
\def\narrowminussub{\kern -.02truein - \kern -.01truein}

\def\kh{K\"{a}hler}

\def\b{{\beta}}
\def\a{{\alpha}}
\def\g{{\gamma}}
\def\e{{\epsilon}}

\def\r{{\rightarrow}}

\def\frac#1#2{{#1\over #2}}

\def\com#1#2{{ \left[ #1, #2 \right] }}

\def\acom#1#2{{ \left\{ #1, #2 \right\} }}

\def\IZ{\relax\ifmmode\mathchoice
{\hbox{\cmss Z\kern-.4em Z}}{\hbox{\cmss Z\kern-.4em Z}}
{\lower.9pt\hbox{\cmsss Z\kern-.4em Z}}
{\lower1.2pt\hbox{\cmsss Z\kern-.4em Z}}\else{\cmss Z\kern-.4em
Z}\fi}
\def\IB{\relax{\rm I\kern-.18em B}}
\def\IC{{\relax\hbox{$\inbar\kern-.3em{\rm C}$}}}
\def\ID{\relax{\rm I\kern-.18em D}}
\def\IE{\relax{\rm I\kern-.18em E}}
\def\IF{\relax{\rm I\kern-.18em F}}
\def\IG{\relax\hbox{$\inbar\kern-.3em{\rm G}$}}
\def\IGa{\relax\hbox{${\rm I}\kern-.18em\Gamma$}}
\def\IH{\relax{\rm I\kern-.18em H}}
\def\II{\relax{\rm I\kern-.18em I}}
\def\IK{\relax{\rm I\kern-.18em K}}
\def\IP{\relax{\rm I\kern-.18em P}}

\def\p{\partial}

\font\cmss=cmss10 \font\cmsss=cmss10 at 7pt
\def\IR{\relax{\rm I\kern-.18em R}}

\def\f{\psi}

\def\s{{\sigma}}
\def\1{{\bf 1}}
\def\3{{\bf 3}}
\def\7{{\bf 7}}
\def\2{{\bf 2}}
\def\8{{\bf 8}}
\def\ec{{\bf 8_c}}

\def\es{{\bf 8_s}}
%

%
%
\def\eqnn#1{\xdef #1{(\secsym\the\meqno)}\writedef{#1\leftbracket#1}%
\global\advance\meqno by1\wrlabeL#1}
\def\eqna#1{\xdef #1##1{\hbox{$(\secsym\the\meqno##1)$}}
\writedef{#1\numbersign1\leftbracket#1{\numbersign1}}%
\global\advance\meqno by1\wrlabeL{#1$\{\}$}}
\def\eqn#1#2{\xdef #1{(\secsym\the\meqno)}\writedef{#1\leftbracket#1}%
\global\advance\meqno by1$$#2\eqno#1\eqlabeL#1$$}



\lref\rsegal{G. Segal and A. Selby, \cmp{177}{1996}{775}. }
\lref\rdorold{N. Dorey, V. Khoze, M. Mattis, D. Tong and S. Vandoren, 
hep-th/9703228, \np{502}{1997}{59}.}
\lref\rsen{A. Sen, hep-th/9402002, Int. J. Mod. Phys. {\bf A9} (1994) 3707;
hep-th/9402032, \pl{329}{1994}{217}. }
\lref\rcallias{C. Callias, \cmp{62}{1978}{213}\semi E. Weinberg, 
\prd{20}{1979}{936}.}
\lref\rblum{J. Blum, hep-th/9401133, \pl{333}{1994}{92}.}
\lref\rsgreen{M. B. Green and S. Sethi, hep-th/9808061.}
\lref\rbateman{H. Bateman, ed. A. Erdelyi, 
{\it Higher Transcendental Functions, vol. 2}, McGraw-Hill Book Company, 1953. }
\lref\ryi{P. Yi, hep-th/9704098, \np{505}{1997}{307}.}
\lref\rsavmark{S. Sethi and M. Stern, hep-th/9705046, \cmp{194}{1998}{675}.}
\lref\rgg{M. B. Green and M. Gutperle, hep-th/9711107, JHEP 01 (1998) 05.}
\lref\rgreg{G. Moore, N. Nekrasov and S. Shatashvili, hep-th/9803265.}
\lref\rSS{S. Sethi and L. Susskind, hep-th/9702101,
    \pl{400}{1997}{265}.}
\lref\rBS{T. Banks and N. Seiberg,
    hep-th/9702187, \np{497}{1997}{41}.}
\lref\rreview{N. Seiberg, hep-th/9705117.}
\lref\rpp{J. Polchinski and P. Pouliot, hep-th/9704029, \prd{56}{1997}{6601}.}
\lref\rds{M. Dine and N. Seiberg, hep-th/9705057, \pl{409}{1997}{239}.}
\lref\rdorey{N. Dorey, V. Khoze and M. Mattis, hep-th/9704197, 
\np{502}{1997}{94}.}
\lref\rbf{T. Banks, W. Fischler, N. Seiberg and L. Susskind, hep-th/9705190, 
\pl{408}{1997}{111}.}

\lref\rK{N. Ishibashi, H. Kawai, Y. Kitazawa and A. Tsuchiya, hep-th/9612115.}
\lref\rCallias{C. Callias, Commun. Math. Phys. {\bf 62} (1978), 213.}
\lref\rPD{J. Polchinski, hep-th/9510017, \prl{\bf 75}{1995}{47}.}
\lref\rWDB{E. Witten,  hep-th/9510135, Nucl. Phys. {\bf B460} (1996) 335.}
\lref\rSSZ{S. Sethi, M. Stern, and E. Zaslow, Nucl. Phys. {\bf B457} (1995)
484.}
\lref\rGH{J. Gauntlett and J. Harvey, Nucl. Phys. {\bf B463} 287. }
\lref\rAS{A. Sen, Phys. Rev. {\bf D53} (1996) 2874; Phys. Rev. {\bf D54} (1996)
2964.}
\lref\rWI{E. Witten, Nucl. Phys. {\bf B202} (1982) 253.}
\lref\rPKT{P. K. Townsend, Phys. Lett. {\bf B350} (1995) 184.}
\lref\rWSD{E. Witten, Nucl. Phys. {\bf B443} (1995) 85.}
\lref\rASS{A. Strominger, Nucl. Phys. {\bf B451} (1995) 96.}
\lref\rBSV{M. Bershadsky, V. Sadov, and C. Vafa, Nucl. Phys. {\bf B463}
(1996) 420.}
\lref\rBSS{L. Brink, J. H. Schwarz and J. Scherk, Nucl. Phys. {\bf B121}
(1977) 77.}
\lref\rCH{M. Claudson and M. Halpern, Nucl. Phys. {\bf B250} (1985) 689.}
\lref\rSM{B. Simon, Ann. Phys. {\bf 146} (1983), 209.}
\lref\rGJ{J. Glimm and A. Jaffe, {\sl Quantum Physics, A Functional Integral
Point of View},
Springer-Verlag (New York), 1981.}
\lref\rADD{ U. H. Danielsson, G. Ferretti, B. Sundborg, Int. J. Mod. Phys. {\bf
A11} (1996) 5463\semi   D. Kabat and P. Pouliot, Phys. Rev. Lett. {\bf 77}
(1996), 1004.}
\lref\rDKPS{ M. R. Douglas, D. Kabat, P. Pouliot and S. Shenker,
hep-th/9608024,
Nucl. Phys. {\bf B485} (1997), 85.}
\lref\rhmon{S. Sethi and M. Stern, Phys. Lett. {\bf B398} (1997), 47.}
\lref\rBFSS{T. Banks, W. Fischler, S. H. Shenker, and L. Susskind,
Phys. Rev. {\bf D55} (1997) 5112.}
\lref\rBHN{ B. de Wit, J. Hoppe and H. Nicolai, Nucl. Phys. {\bf B305}
(1988), 545\semi
B. de Wit, M. M. Luscher, and H. Nicolai, Nucl. Phys. {\bf B320} (1989),
135\semi
B. de Wit, V. Marquard, and H. Nicolai, Comm. Math. Phys. {\bf 128} (1990),
39.}
\lref\rT{ P. Townsend, Phys. Lett. {\bf B373} (1996) 68.}
\lref\rLS{L. Susskind, hep-th/9704080.}
\lref\rFH{J. Frohlich and J. Hoppe, hep-th/9701119.}
\lref\rAg{S. Agmon, {\it Lectures on Exponential Decay of Solutions of
Second-Order Elliptic Equations}, Princeton University Press (Princeton) 1982.}
\lref\rY{P. Yi, hep-th/9704098.}
\lref\rDLhet{ D. Lowe, hep-th/9704041.}
\lref\rqm{M. Claudson and M. Halpern, \np{250}{1985}{689}\semi
R. Flume, Ann. Phys. {\bf 164} (1985) 189\semi
M. Baake, P. Reinecke and V. Rittenberg, J. Math. Phys. {\bf 26} (1985) 1070.}
\lref\rbb{K. Becker and M. Becker, hep-th/9705091, \np{506}{1997}{48}\semi
K. Becker, M. Becker, J. Polchinski and A. Tseytlin, hep-th/9706072,
\prd{56}{1997}{3174}.}
\lref\rss{S. Sethi and M. Stern, hep-th/9705046. }
\lref\rpw{J. Plefka and A. Waldron, hep-th/9710104, \np{512}{1998}{460}.}
\lref\rhs{M. Halpern and C. Schwartz, hep-th/9712133.}
\lref\rlimit{N. Seiberg hep-th/9710009, \prl{79}{1997}{3577}\semi
A. Sen, hep-th/9709220.}
\lref\rentin{D.-E. Diaconescu and R. Entin, hep-th/9706059,
\prd{56}{1997}{8045}.}
\lref\rgreen{M. B. Green and M. Gutperle, hep-th/9701093, \np{498}{1997}{195}.}
\lref\rpioline{B. Pioline, hep-th/9804023.}
\lref\rgl{O. Ganor and L. Motl, hep-th/9803108.}
\lref\rds{M. Dine and N. Seiberg, hep-th/9705057, \pl{409}{1997}{209}.}
\lref\rberg{E. Bergshoeff, M. Rakowski and E. Sezgin, \pl{185}{1987}{371}.}
\lref\rBHP{M. Barrio, R. Helling and G. Polhemus, hep-th/9801189.}
\lref\rper{P. Berglund and D. Minic, hep-th/9708063, \pl{415}{1997}{122}.}
\lref\rspin{P. Kraus, hep-th/9709199, \pl{419}{1998}{73}\semi
J. Harvey, hep-th/9706039\semi
J. Morales, C. Scrucca and M. Serone, hep-th/9709063, \pl{417}{1998}{233}.}
\lref\rdine{M. Dine, R. Echols and J. Gray, hep-th/9805007.}
\lref\rber{D. Berenstein and R. Corrado, hep-th/9702108, \pl{406}{1997}{37}.}
\lref\rnonpert{A. Sen, hep-th/9402032, \pl{329}{1994}{217}; hep-th/9402002,  
Int. J. Mod. Phys. {\bf
A9} (1994) 3707\semi
N. Seiberg and E. Witten, hep-th/9408099, \np{431}{1995}{484}.
}
\lref\rpss{S. Paban, S. Sethi and M. Stern, hep-th/9805018.}
\lref\rpsst{S. Paban, S. Sethi and M. Stern, hep-th/9806028. }
\lref\rperiwal{V. Periwal and R. von Unge, hep-th/9801121.}
\lref\rfer{M. Fabbrichesi, G. Ferreti and R. Iengo, hep-th/9806018.}

\Title{\vbox{\hbox{hep-th/9808119}
\hbox{DUK-CGTP-98-09, IASSNS--HEP--98/71, UTTG-11-98}}}
{\vbox{\centerline{Summing Up Instantons in Three-Dimensional}
\vskip8pt\centerline{Yang-Mills Theories}}}
\centerline{Sonia Paban$^\ast$\footnote{$^1$} {paban@zippy.ph.utexas.edu}, 
Savdeep
Sethi$^\spadesuit$\footnote{$^2$} {sethi@sns.ias.edu} and Mark
Stern$^\dagger$\footnote{$^3$} {stern@math.duke.edu} }
\medskip\centerline{ $\ast$ \it Theory Group, Department of Physics,
University of Texas, Austin, TX 78712, USA}
\medskip\centerline{$\spadesuit$ \it School of Natural Sciences, Institute for
Advanced Study, Princeton, NJ 08540, USA}
\medskip\centerline{$\dagger$ \it Department of Mathematics, Duke University,  
Durham, NC 27706, USA}

\vskip 0.5in

We show that the four derivative terms in the effective action of 
three-dimensional N=8 Yang-Mills theory are determined by supersymmetry. 
These terms receive both perturbative and non-perturbative corrections. 
Using our technique for constraining the effective action, we are able to 
determine the exact form of the eight fermion terms in the supersymmetric 
completion of the $F^4$ term, including all instanton corrections. As a 
consequence, we argue that the integral of the Euler density over $k$ 
monopole moduli space in $SU(2)$ Yang-Mills is determined by our 
non-renormalization theorem for all values of $k$.

\vskip 0.1in
\Date{8/98}

\newsec{Introduction}

In three dimensions unlike other dimensions, the effective action of  
Yang-Mills theory
with sixteen supersymmetries is expected to receive non-perturbative  
corrections at low orders
in a derivative expansion. Let us recall that for $U(N)$ gauge theory, the  
effective action
at generic points on the Coulomb branch is described by a sigma model with   
$7N$ scalars
$\phi^i$ where $i=1,\ldots, 7$. There are also $N$ additional scalars obtained 
by 
dualizing the $N$ abelian gauge fields. We will only consider the effective  
action which
is local in terms of the dual scalars, rather than local in terms of the  
gauge-fields. The
coupling constant in the theory, $g^2$, has mass dimension one. Therefore to  
study the
infra-red theory, we take the coupling $g^2 \r \infty$. The metric on the  
moduli space is flat
and if we take canonical kinetic terms for all fields including the dual  
scalars $ \s$,
$$ L  = {1\over g^2} \int{d^3x \,  {1\over 2} \p_\mu \s \p^\mu \s + \sum_i  
{1\over 2}
\p_\mu \phi^i \p^\mu \phi^i,}$$
then the dual scalars are periodic with a period proportional to $ g^2$. The  
moduli space is
then $ S^{N}( \IR^7\times S^1). $  At the origin of the moduli space,
the theory flows to an interacting $Spin(8)$ invariant fixed point in the  
infra-red \refs{\rSS,
\rBS,\rreview}. The $Spin(8)$ invariance is crucial if matrix theory \rBFSS\  
is to correctly
describe the light-cone quantized type IIB string. For the most part, we  
will consider the
$SU(2)$ theory.

While the metric
on the moduli space of the theory is flat, the higher derivative terms in  
the effective action
must receive non-perturbative corrections, or the effective action
will fail to be $Spin(8)$ invariant in the strong coupling limit. The one instanton
contribution to the $F^4$ terms was first computed in \rpp. Subsequently,  
the computation
was extended to multi-instanton effects in \rdorey. An independent argument for the
existence of instanton contributions to the $F^4$ terms was presented in \rds. 
The importance of the instanton corrections  
for the matrix
description of the type IIB string
was discussed in \rbf. The aim of this paper is to show that the same  
technique used
in \refs{\rpss, \rpsst} easily extends to this case as briefly mentioned in  
\rpss. We will be
able to completely determine the form of the four derivative terms in the  
effective action, including all perturbative
and non-perturbative effects. 

For the four derivative terms, the various contributions to the effective  
action take the form
of a one-loop contribution
together with perturbative corrections around multi-instanton configurations. 
It is quite remarkable that the instanton corrections are completely captured
by the solution of a family of first order differential equations. Relations 
of a similar flavor have appeared in numerous places in both string theory 
and field theory.

By matching the leading instanton contributions to semi-classical computations
\rdorey,
we are able to extract the integral of the Euler density over $k$ monopole 
moduli
space for all $k$. This is completely analogous to the way that the bulk term 
for the 
$L^2$ index for $k$ D0-branes \refs{\rsavmark, \ryi} was extracted from the 
coefficient 
of the ${\cal R}^4$ term in type IIB supergravity \rgg. The coefficient of the 
${\cal R}^4$ term is also determined by supersymmetry \rsgreen. In that case, a 
proof of 
the 
conjectured relation was provided by \rgreg. In this case involving monopoles, 
the
predicted data is more geometrical since it involves the behavior of the metric
on $k$ monopole moduli space.\foot{For non-compact  manifolds like the 
monopole
moduli spaces, the integral of the Euler density is actually weakly metric 
dependent rather than
being simply topological. For this reason, what is usually called the Euler 
index 
need not be integer.
In the case of D0-branes, as an analogous example, the bulk contribution to the 
$L^2$ index is actually not integer.}
This fascinating relation between the properties of various soliton 
configurations and supersymmetry
constraints cries out for a deeper explanation. Lastly, we should  
stress that the same techniques used for $F^4$ can be used to determine 
the $F^6$ terms in three dimensions  as in \rpsst.

\newsec{Determining the Four Derivative Terms in Three Dimensions}
\subsec{An equation for the eight fermion terms}

For simplicity, let us use $\phi^8$ to denote the dual scalar $\s$. For any 
finite
coupling, the symmetry group in three dimensions is $SO(2,1)\times Spin(7)$, 
where 
$SO(2,1) \sim SL(2, \IR)$ is the Lorentz group. The scalars $\phi^i$ where  
$i=1,\ldots, 7$
transform in the $(\1,\7)$ of $SO(2,1)\times Spin(7)$. There are also  
fermions $\f_{\a a}$
which transform in the $(\2,\8 )$ of the symmetry group, where $\a=1,2$ and  
$a=1,\ldots,8$.

While the manifest symmetry group of the Lagrangian is only $Spin(7)$, the  
lowest order
supersymmetry transformations actually exhibit the full infra-red $Spin(8)$  
invariance. We will shift between $Spin(7)$ and $Spin(8)$ labels frequently. In 
the
infra-red, the dual scalar $\phi^8$ combines with the seven scalars $\phi^i$  
to give a vector
$\phi^m$ of $Spin(8)$ where $m=1,\ldots, 8$. The fermions lift to the $(\2,\ec 
)$ 
representation of $SO(2,1)\times Spin(8)$.
The supersymmetry transformations then take the form,
\eqn\susylow{ \eqalign{ \delta \phi^m &=  -i \e^\a_{\dot b} \, \g^m_{{\dot b}  
a} \, \f_{\a a} \cr
\delta  \f_{\a a} &=  \e^\b_{\dot b} \, \g^m_{{\dot b} a}\, \tau^\mu_{\b \a}  
\p_\mu \phi^m, }}
where we have introduced a Grassmann parameter $\e^\a_{\dot a}$ transforming  
in the $(\2, \es)$
of the symmetry group. An explicit realization for the $\tau^\mu$ where 
$\mu=0,1,2$ can 
be given in terms of the Pauli matrices, 
$$ \tau^0=-i \sigma_2, \quad \tau^1= \sigma_3, \quad \tau^2= \sigma_1. $$
With spacetime signature $(-,+,+)$, the $\tau^\mu$ satisfy the relations: 
\eqn\tt{ \eqalign { \acom{\tau^{\mu}}{\tau^{\nu}} & = 2 \eta^{\mu\nu}, \cr 
\com{\tau^\mu}{\tau^\nu} &= 2 \e^{\mu \nu \lambda} \tau_{\lambda}. \cr }}
We will
actually only need to know the variation of the bosons $\phi^m$. When the  
$F^4$ terms are included,
there are higher derivative
corrections to both $\delta \phi$ and $\delta \f$ in \susylow\  but
as in \rpss, they will not play a role in determining the four derivative  
terms.

As in the case of lower dimensional theories, we will focus on the eight  
fermion terms in the
supersymmetric
completion of the $F^4$ term. If we schematically denote the eight fermion  
terms by,
\eqn\eight{ f^{(8)}( \phi) \f^8, }
then it is easy to see that the variation of \eight\ in the Lagrangian using  
\susylow\
contains a piece
with seven fermions and a piece with nine fermions. The same argument given  
in \rpss\ allows
us to conclude that the variation of no other term in the Lagrangian  
contains nine fermions.
Therefore, the piece with nine fermions in the variation of \eight\ must  
vanish giving the
equation,
\eqn\constraint{ \g^m_{{\dot b} a} \, \f_{\a a} {\partial \over \partial \phi^m}
\left( f^{(8)} (\phi) \f^8 \right) =0, }
for every $ {\dot b}$ and $\a$. To proceed, we need to know something
about the possible eight fermion terms.

\subsec{The basic fermion bilinears}

The eight fermion terms contain no spacetime derivatives and must be  
invariant under the
Lorentz group $SO(2,1)$. They can be built from a few fundamental forms  
bilinear in the
fermions. To count the number of basic bilinears, we note that
\eqn\group{ \left[ \, (\2, \8) \otimes (\2, \8) \,  \right]_a =  (\3, \7)\oplus
 ( \3, {\bf 21}) \oplus (\1,\1)\oplus (\1, {\bf 35}), }
as representations of $SO(2,1)\times Spin(7)$. If we let $ \g^i$ be a basis  
for the $Spin(7)$
Clifford algebra then we can define,\foot{When we want to view these $Spin(7)$
gamma matrices as part of the $ \ec $ or $\es $ representations of $Spin(8)$, 
we will use dotted indices; for example, $\g_{\dot{a}b}$.}
\eqn\defs{ \eqalign{ \g^{ij} &= {1\over 2!} ( \g^i \g^j - \g^j \g^i) \cr
\g^{ijk} &= {1\over 3!}( \g^i \g^j \g^k - \g^j \g^i \g^k + \ldots). \cr}}
The antisymmetric matrices $ \{ \g^i, \g^{ij} \}$ and the symmetric matrices  
$ \{ 1, \g^{ijk} \}$
constitute a basis out of which we can construct all operators involving  
fermions. The four
basic bilinears appearing in \group\ are easily found,
\eqn\basic{ \bar{\f} \tau^\mu \g^i \f, \quad \bar{\f}  \tau^\mu \g^{ij} \f, 
\quad  
\bar{\f} \f,
\quad  \bar{\f} \g^{ijk} \f_ ,}
where $\bar{\f}=\f^T \tau_0$. Using various Fierz identities,  
we could in principle
determine all possible eight fermion structures. Instead, we will take a  
different route to
fix all possible eight fermion terms. The only observation we need to make  
concerns the spacetime
indices. To get a Lorentz invariant structure, we will eventually need to  
contract all $\mu$
indices. However, because of the following Fierz identity, 
$$ (\tau^{\mu})_{\a \b}  (\tau_{\mu})_{\s \delta} = \delta_{\a \b} \delta_{\s 
\delta}
- 2 \delta_{\a \delta} 
\delta_{\s \b}, $$
we actually only need to consider the last two structures in \basic. This  
allows us to observe that
any pure fermion structure must contain equal numbers of $\f_{1a}$ and $  
\f_{2a}$. 

There is a second way to see this condition. Let $T$ denote a diagonal element
of the Lorentz group $SL(2,\IR)$ with eigenvalues $t>1$ and $t^{-1}$. Then we 
may pick a basis such that, 
$$T\f_{1a} = t\f_{1a}, $$ 
and 
$$T\f_{2a} = t^{-1}\f_{2a}.$$
Clearly then an eight fermion term which is invariant under $T$ 
must have four $\f_{1a}$  and four $\f_{2a}$ fermions.  We can now 
turn to the  general form of the
eight fermion terms.

\subsec{Classifying the possible eight fermion terms}

First note that the eight fermion structures that can be constructed from  
the last two basic
bilinears \basic\ can be contracted with at most four scalar fields. Further  
$\phi^8$ can
only appear in the form $e^{ in \phi^8 / g^2}$ since there are no spacetime  
derivatives.
We will normalize $ \phi^8$ to have period $2\pi g^2$. We can then classify  
eight fermion
terms in the following way,
\eqn\list{ T_{m_0}, \quad  \phi^i T^{i}_{m_1}, \quad
\phi^i \phi^j T^{ij}_{m_2}, \quad \phi^i \phi^j \phi^k T^{ijk}_{m_3},
\quad \phi^i \phi^j \phi^k \phi^l T^{ijkl}_{m_4}, }
where $T$ is an eight fermion structure and $i,j,k,l<8$. The label $m_k=1, 
\ldots, N_k$ and  $N_k$ is the number of independent structures that contain 
$k$ scalars.
For example, the possible eight fermion terms with four scalars can take the 
form, 
$$ \sum_{m=1}^{N_4} f^{4, m}_{n}(r) e^{in \phi^8 / g^2} \left( \phi^i \phi^j 
\phi^k  
\phi^l
T^{ijkl}_{m} \right), $$
under the assumption that there are any such four scalar structures. Here we 
have 
defined $r$ to be $ \sqrt{ \phi^i \phi^i}$.
Now as in  \refs{\rpss, \rpsst}, 
we can obtain a weaker condition than \constraint\ by applying 
the operator, 
$$\g^n_{c {\dot b}} \, {d \over d\f_{\a c}} \partial_{\phi^n},$$
to \constraint. After summing over $ {\dot b}$, we obtain the condition: 
\eqn\weaker{ \Delta \left( f^{(8)} (\phi) \f^8 \right) =0,}
where the Laplacian $\Delta$ includes $\phi^8$. 

Let us start by considering the restrictions imposed by the weaker  
constraint \weaker. First note that  the Laplacian commutes with the 
Fourier expansion so we obtain an equation on each Fourier coefficient.
Also note that \weaker\ does
not mix structures with an odd number of scalars and structures with an
even number of scalars. This is not true of the stronger condition \constraint.
Let us start by writing the first of three coupled differential equations from 
\weaker\ for the even scalar structures:

\eqn\evenone{ \eqalign{  \left( {d^2 \over dr^2 } + {14\over r} 
{d\over dr}  - {n^2\over g^4} \right) f^{4,m}_n =0. \cr }}
This gives a standard differential equation for each $m$ and $n$. The 
perturbative contribution corresponds to $n=0$ and the general
solution is,
\eqn\sone{ f^{4,m}_0 = {c_0\over r^{13}} + d_0.}
We can set $d_0=0$ since we demand a solution that decays as $r\r\infty$.  
The remaining solution
is then precisely the one-loop contribution and corresponds to a ${1\over  
r^5} F^4$ term. The
coefficient $c_0$ is then fixed by a one-loop computation. We can conclude that 
if 
there are any non-vanishing four scalar structures then they receive only a
one-loop contribution. Before analyzing the non-perturbative terms with $n \neq 
0$,
let us consider the remaining equations. 

The second equation mixes the two scalar and four scalar structures:
\eqn\eventwo{ \eqalign{
\sum_{p=1}^{N_2} \left( {d^2 \over dr^2 } + {10\over r} 
{d\over dr}  - {n^2\over g^4} \right) f^{2,p}_n  e^{in \phi^8 / g^2} 
\left( \phi^i \phi^j T^{ij}_{p} \right) &=  \cr -12 \sum_{m=1}^{N_4}  f^{4,m}_n  
e^{in \phi^8 / g^2} 
\left( \phi^i \phi^j \delta_{kl} T^{ijkl}_{m} \right). \cr }}
Again we will first consider the perturbative contributions with $n=0$. There
are two possible scenarios: either there is a $  \sim 1/r^{13}$ source term
for the differential equation for a two scalar structure labelled by $p$, or 
there is no source term. In the latter case, the solution takes the form
$$ f^{2,p}_0 \sim {1\over r^9}, $$
but this corresponds to a {\it negative} power of the coupling. All such
solutions are unphysical. The only possible physical (perturbative) two
scalar structures must appear with source terms. They therefore
descend naturally from four scalar structures and always have the form, 
$$   \phi^i \phi^j \delta_{kl} T^{ijkl}_{m}, $$
for some four scalar structure. In this case, we find that
\eqn\stwo{ f^{2,p}_0 = -{6 \over 11} {c_0\over r^{11}},}
which is again one-loop exact. 

The last equation for the even structures mixes the no scalar and 
two scalar structures: 
\eqn\eventhree{ \eqalign{
\sum_{q=1}^{N_2} \left( {d^2 \over dr^2 } + {6\over r} 
{d\over dr}  - {n^2\over g^4} \right) f^{0,q}_n  e^{in \phi^8 / g^2} 
T_{q}  &=  -2 \sum_{p=1}^{N_2}  f^{2,p}_n  
e^{in \phi^8 / g^2} \left( \delta_{ij} T^{ij}_{p} \right). \cr }}
Without a source term, we find that
$$  f^{0,q}_0 \sim {1\over r^5}, $$
which requires negative powers of the coupling. We therefore need to consider 
only zero scalar structures that descend from four scalar structures, and
we find that
\eqn\sthree{ f^{0,q}_0 = {1 \over 33} {c_0\over r^{9}},}
which is again one-loop exact. The even structures are therefore completely
determined at the perturbative level by classifying the four fermion structures.

Let us turn to the odd scalar structures. The first equation constrains the 
three scalar structures alone:
\eqn\oddone{ \eqalign{  \left( {d^2 \over dr^2 } + {12\over r} 
{d\over dr}  - {n^2\over g^4} \right) f^{3,m}_n =0. \cr }}
When $n=0$, the solution takes the form
$$ f^{3,m}_0 \sim {1\over r^{11}}, $$
which requires no power of the coupling. Since the structure cannot appear at
tree-level, there are no
perturbative three scalar structures. The final equation relates the three and 
one
scalar structures:
\eqn\oddtwo{ \eqalign{
\sum_{p=1}^{N_1} \left( {d^2 \over dr^2 } + {8\over r} 
{d\over dr}  - {n^2\over g^4} \right)(f^{1,p}_n)  e^{in \phi^8 / g^2} 
\left( \phi^i T^{i}_{p} \right) &=  \cr -6 \sum_{m=1}^{N_3}  f^{3,m}_n  
e^{in \phi^8 / g^2} 
\left( \phi^i \delta_{jk} T^{ijk}_{m} \right). \cr }}
When $n=0$, the solution is of the form 
$$ f^{1,p}_0 \sim {1\over r^{7}}, $$
which requires a negative power of the coupling constant. Therefore, there are 
no
possible odd scalar structures at the perturbative level.\foot{We wish to thank 
N. Seiberg for pointing
out the following simple argument that excludes odd scalar structures at the 
perturbative level. The Weyl group action given in $(3.2)$ inverts all the scalar 
fields
including $\phi^8$. Since $\phi^8$ does not appear perturbatively, this symmetry
excludes any odd scalar structure.}
Since we know there is
a non-vanishing one-loop contribution to the $F^4$ terms, we can also conclude 
that there must be at least one possible four scalar structure. 

\subsec{Non-perturbative solutions}

At the perturbative level, we have demonstrated that the $F^4$ terms 
receive only one-loop contributions just by using the weak constraint \weaker. 
We
can now consider the non-perturbative solutions where $ n \neq 0$. Consider
the equation, 
\eqn\nonpert{ \left( {d^2 \over dr^2 } + {2\nu+1\over r} 
{d\over dr}  - {n^2\over g^4} \right) f_n(r)=0. }
It is convenient to replace $f_n(r)$ by $ r^{\nu} f_n(r)$, which takes the 
equation to the form:
\eqn\newnonpert{ \left( {d^2 \over dr^2 } + {1\over r} 
{d\over dr}  - {n^2\over g^4} - {\nu^2\over r^2} \right) r^\nu f_n(r)=0. }
The solution to this equation is given in terms of $K_\nu (z)$, a modified 
Bessel 
function of the third kind:
\eqn\solve{ f_n(r) = c_n \, n^{1/2} \, {1\over g^{24}} \left( {g^2\over r}\right)^{\nu}\,
 K_{\nu} (nr/g^2).}
Here $c_n$ is a constant yet to be determined and we have inserted powers of 
the coupling needed on dimensional grounds. Most of the time, $\nu$ will be 
half-integer and then $K_\nu (z)$ has the following useful expansion \rbateman, 
\eqn\expansion{ K_{n+1/2}(z) = 
(\pi/2z)^{1/2}e^{-z}\sum_{m=0}^{n}(2z)^{-m}\frac{\Gamma(n+m+1)}{m!\Gamma(n-m+1)}
.}

Before solving the differential equations that we found in the previous section,
some dimensional analysis is in order. It is convenient to rescale the coupling
into the scalar fields and fermions. After this rescaling the fields $\phi$ have
dimension $1/2$ while the fermions have dimension $1$. The kinetic term takes
the form, 
$$ \int{ d^3x \, {1\over 2} \p_\mu \phi \p^\mu \phi, }$$
with no explicit powers of the coupling and with $\phi^8$ included. The one-loop
corrected $F^4$ terms take the form, 
$$  {1\over g }\int{ d^3x \, {1\over r^5} \left( \p_\mu \phi^i \p^\mu 
\phi^i \right)^2, }$$
with a single power of $1/g$. The powers of $g$ are simply fixed on dimensional
grounds since the action is dimensionless. The one-loop corrected effective
action is not scale-invariant. In the limit of low-energies where $g \r \infty$,
the theory must become scale-invariant. All explicit powers of $g$ must
disappear from the action. We will see how the instanton corrections to the 
$F^4$
term combine with the one-loop correction to give a scale and $Spin(8)$ invariant 
result in
the strong coupling limit as suggested in \rbf.

Let us turn to the solution for the four scalar structure equation \evenone. In
this case, $\nu=13/2$ and the solution for the $n$ instanton sector is given by,
\eqn\evenonesolve{ \eqalign{ f_n^{4,m} e^{in \phi^8 /g} & =  c_n \, n^{1/2} \, {1\over 
g^{14}} \,
\left({g\over r} \right)^{13/2} \, K_{13/2} (nr/g) e^{in \phi^8 /g} \cr 
& = c_n \, n^{1/2}\, {1\over g^{14}} \, 
\left( { \pi  \over 2 n} \right)^{1/2}  e^{-{|n|r\over g} + 
{in \phi^8 \over g}}  
\left( {g^7 \over r^7} + \ldots \, \right), }}
which is exactly the form that we would expect for an $n$ instanton sector
contribution. 

When the four scalars are taken into account, we see that the norm of 
the four scalar term goes like  $ \sim r^{-3}e^{-|n|r/g}$ up to lower order
terms. A 
similar analysis shows that the $k$ scalar term can be written as a 
sum of a solution to the homogeneous Laplace equation with leading order 
behavior $ \sim r^{-3}e^{-|n|r/g}$ and a solution to the  
inhomogeneous equation with leading order 
behavior  $ \sim r^{-4}e^{-|n|r}$. Set 
$$\g^r = {1\over r} \phi^j\g^j.$$
Isolating the part of the constraint equation \constraint\ which equates
 terms of order $r^{-3}e^{-|n|r/g}$ gives for nonzero $n$,
\eqn\next{ \left( \g^r_{\dot{a}b} - i \, {\rm sign(n)} \g^{8}_{\dot{a}b} \right) 
\psi_{\alpha b} \, h = 0, }
where $h$ denotes the $O(r^{-3}e^{-|n|r})$ eight fermion terms. This equality 
comes from differentiating only the exponential $e^{-r|n|/g + in\phi^8/g}$ 
in the constraint equation. 
Once again we pass to a weaker but more pliable equation. We apply 
the operator  $\g^r_{c\dot{a}}\partial_{\psi_{\alpha c}}$ to the above 
inequality, 
and sum over $ \dot{a}$ and $\alpha$ to obtain: 
$$\partial_{\psi_{\alpha b}} \left( \psi_{\alpha b} \, h \right) + 
i \, {\rm sign(n)} \g^{r8}_{cb} \psi_{\alpha b}\partial_{\psi_{\alpha c}} \, h 
= 0.$$ 
Using, 
$$\partial_{\psi_{\alpha t}} \left( \psi_{\alpha t} \, h \right) = 8 h,$$ 
we find that, 
\eqn\eigen{ \left( 8 - i\, {\rm sign(n)} \g^{r8}_{st}\psi_{\alpha s}
\partial_{\psi_{\alpha t}} \right) \, h = 0.}
We have already shown that any eight fermion term contains an equal
number of $\psi_{1 a}$ and $\psi_{2a}$. We can therefore build 
$h$ from terms of the form, 
$$ \f_1 \g^{m_1p_1} \f_1\f_1 \g^{m_2p_2} \f_1 
\f_2 \g^{m_3p_3} \f_2\f_2 \g^{m_4p_4} \f_2, $$
where $ 1 \leq m_i, p_i \leq 8$. By including $\g^8$ in $\g^{mp}$, we 
automatically include both antisymmetric basis elements $ \{ \g^i, \g^{ij} 
\}$.   

The operator 
$i\, {\rm sign(n)} \g^{r8}_{bc} \psi_{\alpha b} \partial_{\psi_{\alpha c}}$ acts 
on a bilinear of the form 
$\psi_{\beta}\g^{mp}\psi_{\beta}$ by sending it to,
$$\psi_{\beta}\, i\, {\rm sign(n)} [\g^{r8},\g^{mp}]\psi_{\beta}.$$ 
An easy computation gives eigenvalues of $2,0,-2$ for this operation 
on bilinears. The largest eigenvalue is associated to eigenvectors 
of the form, 
$$\psi_{\beta}(\g^{pr}\pm i \, {\rm sign(n) } \g^{p8}) \psi_{\beta},$$ 
where $p \neq r,8$. Then 
for $n$ positive, the only 
possible way 
to obtain an eigenvalue of 8 for the eight fermion term as required by 
\eigen\ is if the eight fermion term is of the form,\foot{For $n$ negative,
we replace each bilinear $\psi(\g^{mr} - i\g^{m8})\psi$ by 
$\psi(\g^{mr} + i\g^{m8})\psi$. }
\eqn\finalform{ \eqalign{ & a_1 \left[ \psi_1(\g^{pr} - i\g^{p8})\psi_1 
\psi_1(\g^{pr} - 
i\g^{p8})\psi_1 \psi_2(\g^{qr} - i\g^{q8})\psi_2 \psi_2(\g^{qr} - 
i\g^{q8})\psi_2
\right] + \cr & 
a_2 \left[ \psi_1(\g^{pr} - i\g^{p8})\psi_1 \psi_1(\g^{qr} - i\g^{q8})\psi_1
 \psi_2(\g^{pr} - i\g^{p8})\psi_2 \psi_2(\g^{qr} - i\g^{q8})\psi_2 \right],}}
 for constants $a_1$ and $a_2$. Let us use $s(a_1,a_2)$ to 
denote the structure in \finalform\ for convenience. In fact, up to a scalar
factor, there is a unique combination of $a_1$ and $a_2$ allowed. This can be
seen in the following way: equation \next\ implies that,
$$ (I - i \g^{r8})_{as} \psi_{\a s} h = 0. $$
Note that $ (I - i \g^{r8})/2$ is a Hermitian projection operator with the
complementary orthogonal projection given by its conjugate. Therefore, we see that
it has four zero and four non-zero eigenvalues on each chiral subspace since
conjugation preserves chirality. Now for an eight fermion term to be 
annihilated by each of 
the eight non-zero fermions $ (I-i\g^{r8})_{as} \psi_{\a s}$ simply means that it 
can be expressed as a product of these eight fermions. This uniquely fixes the ratio
of $a_1$ and $a_2$ (unless the two structures are the same). 
Observe that in this notation, the scalar structures are mixed. The 
scalars are hidden in the definition of $\g^r.$ The entire eight fermion term 
for the Fourier mode $n \neq 0$ is then
determined by the four scalar solution \evenonesolve.  

Therefore the eight 
fermion
term is uniquely fixed (up to constants). In the sector with 
instanton 
number $n$ positive, $h$ takes 
the form: 
\eqn\newfinal{   c_n \, n^{1/2} \, {1\over g^{14}} \,
\left({g\over r} \right)^{13/2} \, r^4  \, e^{-nr/g} \, \left( \pi g\over 2nr 
\right)^{1/2} \, e^{in \phi^8 
/g} \, s(a_1,a_2).}
Therefore, the solution of the homogeneous Laplace equation 
for the four scalar term is given by,  
\eqn\nf{ \eqalign{  c_n \, n^{1/2} \, {1\over g^{14}} \,
\left({g\over r} \right)^{13/2} \, r^4 \, K_{13/2} (nr/g) e^{in \phi^8 /g} 
\big[ & a_1 \psi_1\g^{pr}\psi_1 \psi_1\g^{pr}\psi_1  \psi_2\g^{qr}\psi_2 
\psi_2\g^{qr}\psi_2 + \cr & a_2 \psi_1\g^{pr}\psi_1 \psi_1\g^{qr}\psi_1  
\psi_2\g^{pr}\psi_2 \psi_2\g^{qr}\psi_2 \big].}} 
Similarly the coefficients of the different scalar structures in 
the solution of the homogeneous Laplace equation can simply be extracted from 
\newfinal\ by taking the $k$ scalar component of \newfinal\ 
and replacing $e^{-nr/g}(\pi g/2nr)^{1/2}$ by 
the appropriate Bessel function, $K_{5/2 + k}(rn/g).$ 
These then determine the solutions to the inhomogeneous equations. In this way,
the eight fermion term is completely determined in the sector with
instanton number $n$.
What remains is for us to determine the coefficients $c_n$.

\newsec{ Scale and $Spin(8)$ Invariance}
\subsec{Constraints from gauge invariance}
In the strong coupling limit, the effective action must become scale and 
$Spin(8)$
invariant. The eight fermion term \newfinal\ contains both even and odd 
structures. 
Both are needed for $Spin(8)$ invariance. Schematically, the three scalar 
structure
goes over to the term, 
$$ \sum_n f_n^3(r) \phi^i \phi^j \phi^k e^{in\phi^8/g} T^{ijk} 
\quad\r\quad \phi^i \phi^j 
\phi^k \phi^8 T^{ijk8}, $$
in the strong coupling limit, and similarly for the one scalar term. However, we 
will
not actually need to assume $Spin(8)$ invariance to determine the constants 
$c_n$.
First note that hermiticity of the eight fermion term implies that, 
\eqn\herm{ c_{-n}  = c_n^*.}
Since we are considering the effective action of a full $SU(2)$ Yang-Mills 
theory,
we also need to impose invariance under the residual Weyl group action:
\eqn\weyl{ \phi^i \r - \phi^i, \qquad \phi^8 \r - \phi^8, \qquad \f \r -\f.}
Invariance under the Weyl group then implies that, 
\eqn\hermtwo{ c_{-n} = c_n, }
so the coefficients $c_n$ are real. 

\subsec{Determining the $c_n$ coefficients}

Let us now show that the $ g\r \infty$ limit results in a $Spin(8)$ invariant 
eight fermion term. Alternately, the $c_n$ coefficients can be determined by
assuming the $Spin(8)$ invariance of the strong coupling limit. We will see that
our argument gives results which agree with this second approach, and hence
constitutes an independent argument for $Spin(8)$ invariance. 

So far, we have 
shown that the perturbative eight fermion
term is one-loop exact with the coefficient $c_0$ fixed by the loop computation.
The perturbative eight fermion term is completely determined by specifying the 
four scalar structure. We have also shown that the instanton sector $n$ eight 
fermion term is completely determined up to a  constant $c_n$.
Let us focus on the $g$ dependence of the eight fermion term, which we will
denote $f_g$ where $f_g = f_n (g) e^{in \phi^8/g}$. The full Yang-Mills theory 
is
well-behaved at low-energies since away from the singularities in the moduli 
space,
the theory is infra-red free. We can therefore assume that the limit of the 
eight
fermion term $f_g$ and its first six derivatives exists as $g\r \infty$ and 
converges 
to a limit $F$ and its first six derivatives in the
following sense: the eight fermion term $F$ which describes the strong coupling 
limit
can be made periodic with period $2\pi g$ by averaging, 
\eqn\defed{F_g =  \sum_{j\in \IZ} F(\phi^8+2\pi gj).}
We can express $F_g$ in terms of its Fourier modes $\hat{F} ( n/2\pi g)$, 
\eqn\fourier{F_g =  \sum_{n\in \IZ}{1\over 2\pi g} \hat{F} ( n/ 2\pi g)
e^{in\phi^8/g}.}
Note that we are not assuming that $F$ is $Spin(8)$ invariant. The appropriate 
notion of convergence of $f_g$ to $F$ is that, 
$$ g^k \left[ f_n(g) - {1\over 2\pi g } \hat{F}( n/ 2\pi g) \right], $$  
tends to zero as $g\r\infty$ for $0\leq k \leq 6$. In
fact, if this were not the case, the notion of an effective action would make
little sense.  We will see why we need the first
 six derivatives momentarily. 

On dimensional grounds, we know that $f_n = c_n \, {1\over g^7}\, h_n$
where $h_n$ is determined from \newfinal\ by choosing $h_n$ to be normalized
so that $n$ and $g$ only appear together in the combination $|n|/g$.  
Now the same analysis we applied to $f_g$ must apply equally to 
$F_g$ since it satisfies the same constraint equations. So  
there is an expansion, 
\eqn\expand{{1\over 2\pi }\hat{F}(n/2\pi g) = C_n h_n}
The left-hand-side depends on $n$ and $g$ only in the combination $n/g$ and 
therefore the same is true for the  
right-hand-side. Then $C_n = C(n/g)$ and,
$$ c_n - g^{6}C(n/g) \rightarrow 0.$$
We therefore learn that the first six terms in a Taylor expansion of $C(n/g)$ 
about zero vanish, which implies that
$$c_n - g^6 C^{(6)}(0) \, {1\over 6!} \, 
\left({n\over g}\right)^6 \rightarrow 0 $$
as $g$ tends to infinity. Finally, we can conclude that for $n>0$:
\eqn\coeffs{ c_n = c \, {n^6\over 6!}.}
Note that this formula does not imply that $c_0=0$ because $ h_n$ is singular 
at $n=0$. 
Since we have determined 
the strong coupling limit of the eight fermion term by determining
$c_n$, we can Fourier expand the limiting form and extract the coefficient $c_0$. 
Therefore,
the constant $c$ in \coeffs\ is fixed in terms of $c_0$. For our subsequent
discussion, we will not need to determine this constant explicitly.  

\subsec{A comparision with the $Spin(8)$ result}

We will show by direct computation that  \coeffs\ gives an eight fermion 
term which 
converges to a Spin(8) invariant term in the strong coupling limit. 
More accurately, we will compute the Fourier coefficients of the unique 
Spin(8) invariant term and find agreement with \coeffs.

We now set $F$ to be the following 
harmonic four scalar eight fermion term
\eqn\newF{ \eqalign{\sum_{i,j,m,l\leq 8} {1\over(r^2 + (\phi^8)^2)^{7}}\, 
\phi^i\phi^j\phi^m\phi^l \, \big[ & a_1\psi_1 \g^{pi}\psi_1\psi_1\g^{pj}\psi_1
\psi_2\g^{qm}\psi_2\psi_2\g^{ql}\psi_2 + \cr &
 a_2\psi_1 \g^{pi}\psi_1\psi_1\g^{qj}\psi_1
\psi_2\g^{pm}\psi_2\psi_2\g^{ql}\psi_2 \big].}}
The form of this term is simply fixed on dimensional grounds since the effective
action must be scale-invariant.

We can extract the $c_n$ coefficients from the four scalar part of \newF\ where none of the
four scalars is $\phi^8$. We then need to compute the Fourier transform in 
the $\phi^8$ direction of,  
$$ {1\over (r^2+(\phi^8)^2)^7}.$$
The Fourier component $ \hat F(n/g)$ is given by,
\eqn\newfour{\hat F(n/g) = {2\over 6!} \left({ n\over 2gr} \right)^{13/2} \, \pi^{1/2}
\, K_{13/2}(rn/g).}  
We can now see that the homogeneous four scalar term in $F_g$ is 
given by, 
\eqn\agrees{ \eqalign{ & {1\over \pi g}\sum_{n\in \IZ} {\pi^{1/2} \over 6!} 
K_{13/2}(rn/ g) \, e^{in\phi^8/g} \,
\left({ n\over 2gr} \right)^{13/2} \, r^4 \, 
\big[ a_1 \psi_1\g^{pr}\psi_1\psi_1\g^{pr}\psi_1
 \times \cr & \psi_2\g^{qr}\psi_2\psi_2\g^{qr}\psi_2 + 
a_2 \psi_1\g^{pr}\psi_1\psi_1\g^{qr}\psi_1
 \psi_2\g^{pr}\psi_2\psi_2\g^{qr}\psi_2 \big].} }
The coefficient $C_n$ extracted from \agrees\ agrees with the formula \coeffs.

\newsec{A Comparision With Semi-Classical Computations}

The leading terms in the expansion of \newfinal\ should match semi-classical
instanton computations performed in \refs{\rpp, \rdorey}. For example, 
the leading contribution to the zero scalar structure in \newfinal\ takes
the form,
\eqn\leading{ {1\over r^3} e^{-r+i \phi^8},}
in the one-instanton sector. This agrees with the term in the effective action
computed in \rpp\ up to normalization choices.  

For higher instanton number, the situation is much more interesting. Let us
recall that instantons in three-dimensional gauge theory correspond to 
monopole field configurations in four-dimensional gauge theory. Unlike the case 
of four-dimensional
gauge theory, instanton computations in three dimensions are naturally performed
in the broken gauge theory. Monopole solutions have a scale set by the 
choice of vacuum expectation value so integrals over monopole moduli space do 
not
require the use of cut-off techniques like  constrained instantons used 
in four dimensions. 

This simple observation implies a quite beautiful relation between the geometry
of monopole moduli space and the terms in the effective action that we have
determined. Let us recall that $n$ monopole moduli space ${\cal M}_n $ is a 
$4n$-dimensional hyper\kh\ manifold which has a standard isometric 
decomposition,
$${\cal M}_n = \IR^3 \times {S^1 \times {\cal M}_n^0 \over \IZ_k}.$$
The $\IR^3\times S^1$ parametrize the center of mass and the coordinate 
conjugate
to the electric charge. Quantization of these moduli in four-dimensional N=4 
Yang-Mills leads to a supersymmetric sigma model on $ {\cal M}_n$ described in 
\rblum.
Note that it follows from \rcallias\ that there are $8n$ real 
fermionic zero modes in a charge $n$ monopole background.
These zero modes become superpartners to the bosonic moduli in the 
supersymmetric 
sigma model.  

At first sight, it might seem that the expectation value of an
eight fermion term in the $SU(2)$ Yang-Mills theory would vanish in a sector
with instanton number $ |n|>1$ because of fermion zero mode counting. However,
this is not the case because the moduli for the relative motion of the monopoles 
do
not parametrize a flat space. Rather the semi-classical instanton contribution 
is proportional to the high temperature limit of the twisted partition function
on monopole moduli space,
\eqn\part{ \chi_n = \int_{{\cal 
M}_n^0}{ \, \lim_{\beta \r 0} \, \tr (-1)^F e^{-\beta H} (x,x).}} 
The Hamiltonian is the one describing supersymmetric quantum mechanics on 
$ {\cal M}_n$ \rblum.
This is the bulk contribution to the $L^2$ index which we would want to compute
in sectors with a given electric charge to prove, for example, the conjecture by 
Sen about dyon bound states \rsen. The high temperature limit is a perturbative limit
of the partition function and the term $\chi_n$ reduces to the integral
of the Euler density over $ {\cal M}_n^0$,
\eqn\euler{\chi_n  = \int_{{\cal 
M}_n^0}{ e( {T}{\cal M}_n^0)},}
where $ e(  T {\cal M}_n^0) $ is the Euler class.
If the relative monopole moduli 
space 
${\cal M}_n^0$ had been compact, $\chi_n$ would be the Euler index. 
For $n=2$, the Chern-Gauss-Bonnet theorem can be used to show that 
$\chi_2$ is the topological Euler characteristic $\chi_{n}^{top}$. This seems 
likely 
to remain true for all $n$, but a careful proof of this equality for $n >2$ 
requires a better understanding of the monopole moduli space metric.

Since we 
have determined the exact eight fermion terms from supersymmetry, we should be 
able 
to extract 
$\chi_n$ from our solution by matching the leading semi-classical behavior. This 
seems
quite remarkable.

Fortunately for us, there has been a great deal of interesting work on 
semi-classical 
instanton computations in \refs{\rdorey, \rdorold}. From the analysis of the 
measure
for the $n$ instanton computation performed in \rdorey, we can conclude that:
\eqn\test{ {c_n \over c_1} = n^5  \chi_n.}
However, from \coeffs\ we see that
\eqn\compare{{c_n \over c_1} = n^6, }
and therefore $\chi_n = n$. Remarkably, the results obtained in \rsegal\ have been
used in \rdorey\ to show that 
$\chi_{n}^{top}$ is also $n$. Clearly, a deeper explanation of why supersymmetry 
should determine $\chi_n$ is needed.

\bigbreak\bigskip\bigskip\centerline{{\bf Acknowledgements}}\nobreak
It is our pleasure to thank W. Fischler, D. Kabat and especially N. Seiberg 
for 
helpful conversations.  The
work of S.P. is supported by NSF grant
PHY--9511632, that of S.S. by NSF grant
DMS--9627351 and that of M.S. by NSF grant DMS--9505040 and
NSF grant DMS--9870161.

\vfill\eject

\listrefs
\bye